\documentstyle[aps,epsf,multicol]{revtex}

\begin{document}

\draft

\title{Classification of the pairing transition in finite fermionic systems}
\author{A.~Beli\'c}
\address{Institute of Physics, P.O.B. 57,Belgrade 11001, Yugoslavia}  
\author{ D.~J.~Dean}
\address{Physics Division, Oak Ridge National Laboratory, P.~O.~Box 2008, Oak Ridge, Tennessee 37831}
\author{M.~Hjorth-Jensen} 
\address{Department of Physics, University of Oslo, N-0316 Oslo, Norway}

\maketitle

\begin{abstract}
In this work we employ a simple pairing interaction
model in order to study and classify an 
eventual pairing phase transition in finite
fermionic systems. 
We show that systems with as few as $\sim 10-16$ fermions 
can exhibit clear features reminiscent of a phase transition. 
To classify the nature of the transition we apply two different
numerical methods, one based on standard thermodymanics, and another based on 
a recently  proposed scheme by Borrmann {\em et al.}
[Phys.~Rev.~Lett.~{\bf 84}, 3511 (2000)].
The transition is shown to be of second order, in agreement with 
results for infinite fermionic systems.

\end{abstract}

\pacs{PACS number(s): 05.20.-y, 05.70.Ce, 05.70.Fh, 21.10.Ma}
\begin{multicols}{2}  
The standard BCS theory has been widely used to describe 
systems with pairing correlations and phase transitions to a
superconducting phase for large systems, 
from the solid state to nuclear physics,
with neutron stars as perhaps the largest object in the universe
exhibiting  superfluidity in its interior.  
An eventual superfluid phase in a neutron star will condition 
the neutrino emission and thereby the cooling history of such a
star, in addition to inducing mechanisms such as 
sudden spin ups in the rotational
period of the star; see, for example, Ref.~\cite{hh2000} for a recent review. 
For an infinite system, such as a neutron star, the nature
of the pairing phase transition is well established as second order.
 
When a system of correlated fermions such as electrons or
nucleons is sufficiently small,
the fermionic spectrum becomes
discrete. If the spacing approaches the size of the pairing gap,
superconductivity is expected to break down \cite{anderson59};
however, recent experiments on superconducting ultrasmall aluminum 
grains by Tinkham
{\em et al.} \cite{tinkham9598} revealed the existence of a
spectroscopic gap larger than the average electronic level density.
This feature was interpreted as a reminiscence of superconductivity
and renewed the interest \cite{delft98,mastellone98,sierra99,delft2000}
in studies  of what is the lower size limit for superconductivity.

Other finite fermionic systems such as   
nuclei are expected to
exhibit a variety of interesting phase-transition like phenomena, like
the disappearence of pairing at a critical temperature $T_c\approx
0.5-1$ MeV or the nuclear shape transitions of deformed nuclei
associated with the melting of shell effects at  
$T_c\approx 1-12$ MeV.  
In recent theoretical and experimental studies \cite{yoram2000,andreas2000}
of thermodynamical properties of finite nuclei, the heat capacity 
has been found to exhibit a non-vanishing bump at temperatures proportional to
half the pairing gap. These bumps were interpreted as  signs of the 
quenching of pair correlations, representing in turn features
of the pairing transition for an infinitely large system. 

In the study of phase transitions in e.g., solid state, nuclear and
high energy physics, it is important to know  whether
a given transition really is of first order, discontinuous, 
or if there is a continuous change
in a physical quantity like the mean energy, as in phase transitions
of second order. If one works in the canonical or grand canonical
ensembles, for finite systems
it is rather difficult to decide on the order of the phase
transition. This is due to the fact that in ensembles like the canonical,
any anomaly is smeared over a temperature range of $1/N$, $N$ being the
number of particles. In the analysis of  finite systems, both a 
$\delta$-function peak and a power law singularity sharpen as the
number of particles is increased, making it difficult to distinguish
between the two cases, see, for example, Ref.~\cite{huller}. 

In addition, first order phase transitions in finite systems 
have recently been inferred, theoretically and experimentally,  from 
observed negative heat capacities that 
are associated with anomalous convex intruders in the entropy
versus energy curves, resulting in backbendings in the caloric 
curves; see, for example, 
Refs.~\cite{andreas2000,huller,gross,schmidt01,agostino00,gc00}.
Negative heat capacities are often claimed to appear only in calculations
done in the microcanonical ensemble and are thought to vanish 
in the canonical or grand-canonical ensembles. 

We aim in this Letter to identify the nature of the pairing
transition, if any, in a finite fermionic system. 
Since we are dealing with pairing correlations, our 
Hamiltonian is 
\begin{equation}
   H=\sum_i \varepsilon_i a^{\dagger}_i a_i -G\sum_{ij}
           a^{\dagger}_{i}
     a^{\dagger}_{\bar{\imath}}a_{\bar{\jmath}}a_{j},
     \label{eq:pairHamiltonian}
\end{equation}
where $a^{\dagger}$ and $a$ are fermion creation and annihilation operators, 
respectively. The indices $i$ and $j$ run over the number 
of levels $L$, and the label $\bar{\imath}$ stands for a time-reversed state. 
The parameter $G$ is the strength of the pairing 
force while $\varepsilon_i$ is 
the single-particle energy of level $i$. 

We assume that the single-particle levels are equidistant with a 
fixed spacing $d$.
Moreover, in our simple model, the degeneracy of the single-particle 
levels is set to $2J+1=2$, with $J=1/2$ being the spin of the particle. 
Introducing the pair-creation operator 
$S^+_i=a^{\dagger}_{im}a^{\dagger}_{i-m}$,
one can rewrite the Hamiltonian in Eq.\ (\ref{eq:pairHamiltonian}) as
\begin{equation}
   H=d\sum_iiN_i
     -G\sum_{ij}S^+_iS^-_j,
     \label{eq:pair2}
\end{equation}
where  $N_i=a^{\dagger}_i a_i$
is the number operator.
Seniority $\cal{S}$ 
is a good quantum number and the eigenvalue problem 
can be block-diagonalized
in terms of different seniority values. Loosely speaking, 
the seniority quantum number $\cal{S}$ is equal to the number of 
unpaired particles.

For systems with less than $\sim 16-18$ particles, 
this model can be diagonalized
exactly, and we can obtain {\em all eigenstates}. 
In our studies below, we will always
consider the case of half-filling, i.e., equally many particles
and single-particle levels. This case has the largest dimensionality: for 
16 particles in 16 doubly degenerate single-particle shells, we have a
total of $4\times 10^8$ states. We choose units MeV for the energy and 
set $G=0.2$ MeV in all calculations while we let $d$ vary. 

Through diagonalization 
of the above Hamiltonian
we can define exactly the
density of states $\Omega_N(E)$ for an $N$-particle
system with excitation energy $E$. 
An alternative 
to the exact diagonalization, would be 
to  use Richardson's well-known 
solution \cite{richardson}, 
however, we are interested in {\bf all} eigenstates,
and the amount of numerical labor will most likely be similar.
The density of states is
an essential ingredient in the
evaluation of thermal averages and for the discussion of phase
transitions in finite systems. For nuclei, experimental 
information on the density of states
is expected to reveal important information on nuclear shell
structure, pair correlations and other correlation phenomena
in the nucleonic motion. 

The density of states $\Omega_N(E)$ is the statistical
weight of the given state with excitation energy $E$, and its logarithm
\begin{equation}
      S_N(E)=k_B\ln\Omega_N(E),  \label{eq:entromicro}
\end{equation}
is the entropy (we set Boltzmann's constant
$k_B=1$) of the $N$-particle system.  
The density of states defines also the partition function in the 
microcanonical ensemble and can be used to compute the 
partition function $Z$ of the canonical ensemble through
\begin{equation}
    Z(\beta)=\sum_E\Omega_N(E)e^{-\beta E},
    \label{eq:canonicalpart}
\end{equation}
with $\beta=1/T$ 
the inverse temperature.  With $Z$ it is straightforward to 
generate other thermodynamical
properties such as the mean energy 
$\langle E\rangle$ or the specific heat $C_V$.

The density of states can also be used to define 
the  free energy $F(E)$ in the microcanonical ensemble 
at a fixed temperature $T$ (actually an expectation value in this ensemble), 
\begin{equation}
    F(E)=-T\ln\left[\Omega_N(E)e^{-\beta E}\right]\;.
    \label{eq:freenergy}
\end{equation}
Note that here we include only
configurations at a particular $E$.
 
The above free energy was used by e.g.,
Lee and Kosterlitz \cite{prl90},
based on the histogram approach for studying
phase transitions developed by Ferrenberg and Swendsen \cite{fs88},
in their studies of phase transitions
of classical spin systems. 
If a phase transition is present, a plot of $F(E)$ versus $E$ will show
two local minima which correspond to configurations that are
characteristic of the high and low temperature phases.
At the transition temperature $T_C$ the value of $F(E)$ at the 
two minima equal, while at temperatures below $T_C$, the low-energy
minimum is the absolute minimum. At temperatures above $T_C$, the high-energy
minimum is the largest. If there is no
phase transition, the system developes only one minimum for all temperatures.
Since we are dealing with finite systems, we can study the development 
of the two minima as function of the dimension of the system and thereby 
extract information about the nature of the phase transition. If we are dealing
with a second order phase transition, the behavior of $F(E)$ does not change
dramatically as the size of the system increases. However, if the transition
is first order, the difference in free energy, i.e., 
the distance between the maximum and minimum values, 
will increase with increasing dimension.

To elucidate the nature
of the pairing transition we employ two different approaches,
which both rely on our ability of computing the exact density of states
$\Omega(E)$. 

\begin{figure}
\begin{center}
\setlength{\unitlength}{1mm}
   \begin{picture}(140,220)
   \put(0,0){\epsfxsize=18cm \epsfbox{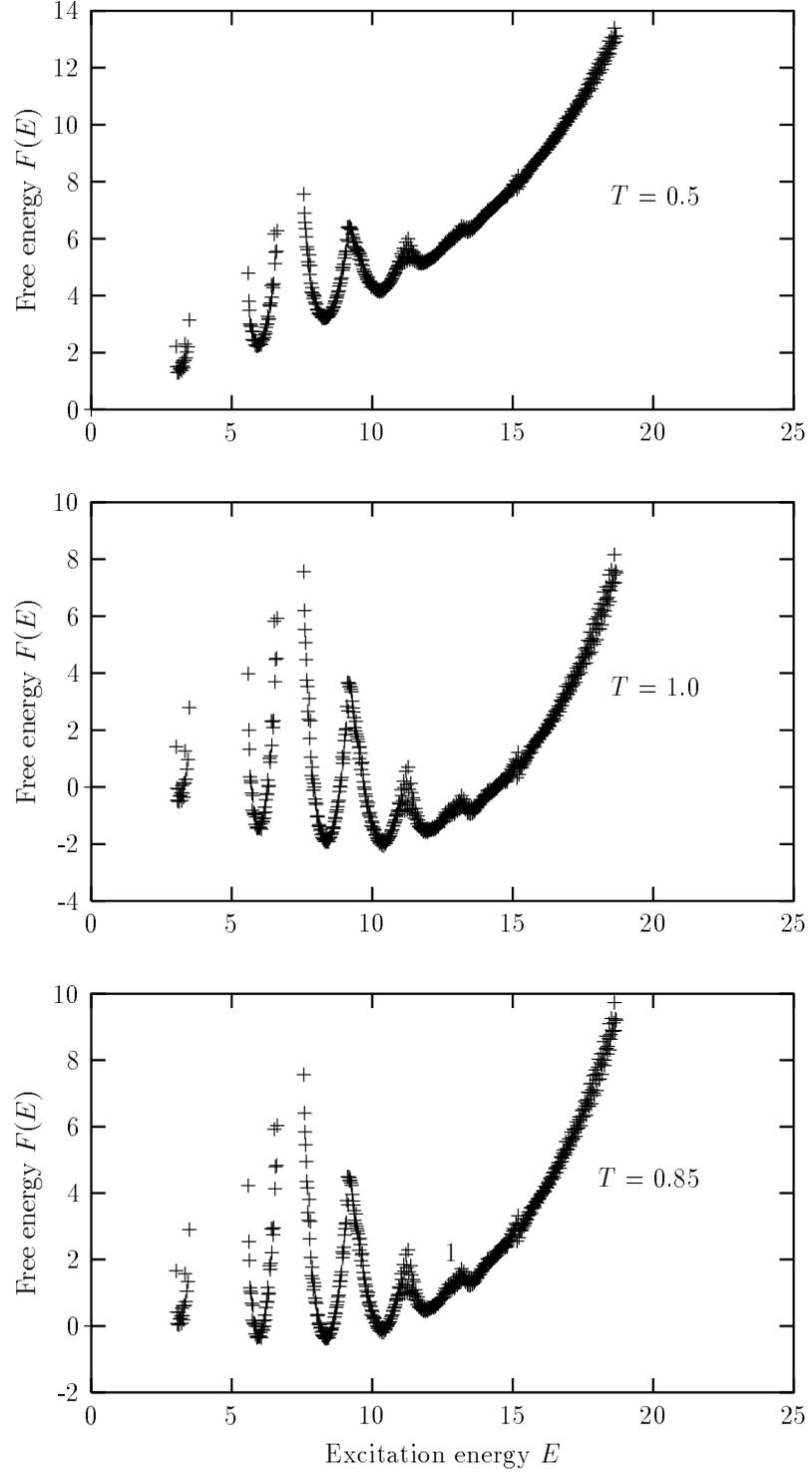}}
   \end{picture}
\end{center} 
\caption{Free energy from Eq.~(\ref{eq:freenergy}) at $T=0.5$, $0.85$ and
         $T=1.0$ MeV  with 
         $d/G=0.5$ with 16 particles in 16 doubly degenerate
         levels. All energies are in units of MeV and 
         an energy bin of $10^{-3}$ MeV has been chosen.}
\label{fig:free_energy16}
\end{figure}

First, we calculate exactly the  
free energy $F(E)$ of Eq.~(\ref{eq:freenergy})
through diagonalization of the pairing Hamiltonian of 
Eq.~(\ref{eq:pairHamiltonian})
for systems with up to 16 particles in $16$ doubly degenerate
levels. 
For $d/G=0.5$ and 16 single-particle levels, 
we develop two clear  minima for the free energy.
This is seen in
Fig.~\ref{fig:free_energy16} where we show the free energy as function of 
excitation energy
using Eq.~(\ref{eq:freenergy}) at temperatures 
$T=0.5$, $T=0.85$ and $T=1.0$~MeV.
The first minimum corresponds to the case where we break one pair.
The second and third minima  correspond
to cases where two and three pairs are broken, respectively. 
When two pairs are broken, corresponding to seniority ${\cal S}=4$, 
the free energy minimum is made up of contributions
from states with ${\cal S}=0,2,4$. These contributions serve to lower
the free energy. 
Similarly, with three pairs
broken we see a new free energy minimum which receives contributions
from ${\cal S}=0,2,4,6$.
At higher excitation energies, population
inversion takes place, and our model is no longer realistic. 

We note that for $T=0.5$~MeV, the minima at lower excitation
energies are favored. 
At $T=1.0$~MeV, the higher energy
phase (more broken pairs) is favored.
We see also, at $T=0.85$~MeV, that 
the free-energy minima where we break two and three pairs 
equal. 
Where two minima coexist, we may have an
indication  of a phase transition. Note however that this is not a 
phase transition in the ordinary thermodynamical sense.
There is no abrupt transition from a purely paired phase to a 
nonpaired phase.  
Instead, our system developes several such intermediate steps
where different numbers of broken pairs can coexist. 
At e.g., $T=0.95$~MeV, we find again two equal minima. For this case,
seniority ${\cal S}=6$ and ${\cal S}=8$ yield two equal minima.
This picture repeats itself for higher seniority and higher temperatures.

If we then focus on the second and third minima, i.e., where we break
two and three pairs, respectively, the difference $\Delta F$ between the 
minimum and the maximum of the free energy, can aid us in distinguishing
between a first order and a second order phase transition. If $\Delta F/N$
remains constant as $N$
increases, we have a second order transition. An increasing $\Delta F/N$
indicates a first order phase transition. 
In Table \ref{tab:free_energy10_16} we display $\Delta F/N$ for 
$N=10$, 12, 14 and 16 at $T=0.85$ MeV. 
It is important to note that the features
seen in Fig.~\ref{fig:free_energy16}, apply to the cases with $N=10$, 12 
and 14 as well, where $T=0.85$~MeV is the temperature where the second and
third minima equal. This means that the temperature where the transition
is meant to take place remains stable as function of number of single-particle
levels and particles. This is in agreement with the simulations of 
Lee and Kosterlitz \cite{prl90}. We find a similar result for the minima
developed at $T=0.95$~MeV, where both ${\cal S}=6$ and ${\cal S}=8$ coexist.
However, due to population inversion, these minima are only seen clearly
for $N=12$, $14$ and $16$ particles.

Table \ref{tab:free_energy10_16} reveals that $\Delta F/N$ is nearly
constant, with  $\Delta F/N\approx 0.5$~MeV, indicating a 
transition of second order. This result is in 
agreement with what is expected for an infinite system. 
It is also easy to see from Fig.~\ref{fig:free_energy16},
that the entropy in the microcanonical ensemble can be convex for 
certain excitation energy ranges, resulting in eventual negative 
heat capacities, as inferred from the 
authors of Refs.~\cite{huller,gross}. The analysis
above however, does not lend
support to interpreting this as a sign of a first order phase transition; see
also the recent work of Moretto {\em et al.} \cite{moretto01}.

We note the important result that for $d/G > 1.5$, 
our free energy, for $N\le 16$, developes
only one minimum for all temperatures. That is, for larger single-particle
spacings, there is no sign of a phase transition. This means that there
is a critical relation between $d$ and $G$ for the appearance of a phase 
transition-like behavior, being a  reminiscence of the thermodynamical limit.
This agrees also with e.g., the results for ultrasmall metallic grains
\cite{delft2000}. 

We next compute the distributions of zeros (DOZ)
of the canonical partition
function and corresponding poles of the specific heat in 
the complex temperature plane, following a 
scheme outlined by Lee and Yang \cite{leeyang}, Grossmann
{\em et al.} \cite{grossmann} and Borrmann {\em et al.} \cite{borrmann2000}. 
Following Ref.~\cite{borrmann2000},
we restrict our discussion to the canonical ensemble and
denote complex temperatures by $ {\cal B} = \beta + i \tau$.
The partition function of Eq.~(\ref{eq:canonicalpart}) is now a function
of ${\cal B}$. 
The authors of Ref.~\cite{borrmann2000} showed that the 
distributions of zeros are able to reveal the
thermodynamic secrets of small systems in a distinct manner.
Major contributions 
to the specific heat come from zeros close to the real axis, and
a zero approaching the real axis infinitely closely causes 
a divergence in the specific heat. 
To characterize the DOZ close to the real axis, one assumes 
that the zeros lie approximately on a straight line.  
Defining the three parameters $\tau$, $\gamma$ and $\alpha$, see
Ref.~\cite{borrmann2000} for details,
one can define entirely the nature of the phase transition.
For a {\sl true} phase
transition in the Ehrenfest sense we have $\tau \rightarrow 0$. For
this case it has been shown \cite{grossmann} that a phase transition is
completely classified by $\alpha$ and $\gamma$.  In the case $\alpha =
0$ and $\gamma = 0$ the specific heat $C_V(\beta)$ exhibits a
$\delta$-peak corresponding to a phase transition of first order. For 
$0 < \alpha < 1$ and $\gamma =0$ (or $\gamma \neq 0$) the transition is of
second order. A higher order transition occurs for  $1 < \alpha $ and
arbitrary $\gamma$.  

\begin{figure}
\begin{center}
   \setlength{\unitlength}{1mm}
   \begin{picture}(140,220)
   \put(30,0){\epsfxsize=9cm \epsfbox{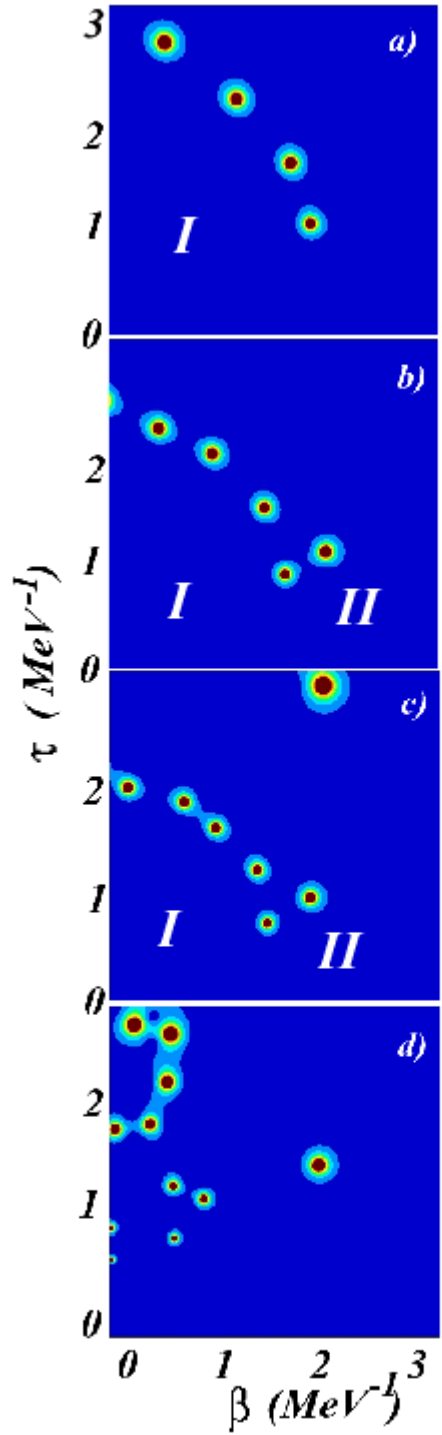}}
   \end{picture}
\end{center} 
\caption{Contour plots of the specific heat in the complex temperature plane
for a) $N=11$, b) $N=14$, and c) $N=16$ particles. Panel d) 
shows the $N=14$ case with weak pairing.  
The spots indicate the locations of the 
zeros of the canonical partition function.} 
\label{fig:contourplot}
\end{figure}

In Fig.~\ref{fig:contourplot} we show contour 
plots of the specific heat $\mid C_v({\cal B})\mid $
in the complex temperature plane
for $N=11$ (a), $14$ (b), and $16$ (c) particles at normal 
pairing $d/G=0.5$ and
the $N=14$ (d) in the weak pairing limit, $d/G=2$. 
The poles are at the center of the dark contour regions.  We see evidence
of two phases in these systems. The first phase, labeled $I$ in 
Fig.~\ref{fig:contourplot}, is a mixed seniority phase while the 
second phase, $II$, is a paired phase with zero seniority and exists
only in even-$N$ systems.  No paired phase exists in 
the $N=11$ system and no clear boundaries are evident 
in the weak pairing case. We find that for (b) and (c) the 
DOZ are apparently distributed along two lines where the 
intersection occurs at $\tau_1$, which is the pole closest to 
the real axis. As the pairing branch (for $\beta >\beta_1$) 
only encompasses two points, we are unable to precisely 
determine $\alpha$ along this branch while $\gamma>0$. Based on
our free energy results discussed above, we believe $\alpha$ 
along this branch will be positive. In the mixed phase branch
(for $\beta<\beta_1$) we find $\gamma<0$, and $\alpha < 0$ in
all normal-pairing cases. The parameter $\tau_1$, which is a measure
of discreteness shows a $\tau_1\sim N^{-1.4 \pm 0.12}$ dependence. 

We note several significant results of this work. From two 
independent methods we find that the transition from the paired seniority
zero ground state to a mixed phase state is second order. The free-energy
analysis also demonstrates that each transition in seniority phases in 
the microcanonical ensemble is of second order. The strength of the
pairing in these systems determines the nature of the phase transitions. 
In particular, for a weakly paired system, we found no evidence 
for two phases, while normal pairing strengths, such as those
found in nuclei, may well exhibit the paired-phase and mixed seniority
phases that we demonstrated in this model. We will include more realistic
interactions to investigate this point in future work. 
We also found, using Auxiliary Field Monte Carlo 
computations for this system \cite{kdl97} together 
with the histogram method of Refs.~\cite{prl90,fs88}, that the energy
fluctuations in the canonical ensemble 
make it rather difficult to extract useful information
on the nature of the phase transitions from these techniques. 

Oak Ridge National Laboratory is
managed by UT-Battelle, LLC for the US Department of Energy 
under Contract No.~DE-AC05-00OR22725. We are indebted to M.~Guttormsen 
and B.~Mottelson for many suggestions and comments.

\end{multicols}

\clearpage

\begin{table}[b]
\begin{tabular}{ccccc}
$N$ & 10 & 12 & 14& 16 \\
$\Delta F/N$ [MeV]   &0.531 & 0.505 & 0.501 & 0.495 \\
\end{tabular} 
\caption{ $\Delta F/N$ for $T=0.85$ MeV. See text for further details.} 
\label{tab:free_energy10_16}
\end{table}

\end{document}